\documentclass[twocolumn,final,balancelastpage,tightenlines,%
                             amsmath,superscriptaddress,a4paper]{revtex4}
                             
\usepackage[dvipdfm]{graphicx}
\usepackage{latexsym}
\begin{document}
\title{%
Low temperature metallic state
induced by electrostatic carrier doping of SrTiO$_3$%
}
\author{H. Nakamura}
\affiliation{Correlated Electron Research Center (CERC),\\
	National Institute of Advanced Industrial Science and Technology (AIST),\\
	Tsukuba 305-8562, Japan}
\affiliation{Department of Advanced Materials,
	University of Tokyo, Kashiwa 277-8581, Japan.}
\author{I. H. Inoue}
\affiliation{Correlated Electron Research Center (CERC),\\
	National Institute of Advanced Industrial Science and Technology (AIST),\\
	Tsukuba 305-8562, Japan}
\author{H. Takagi}
\affiliation{Correlated Electron Research Center (CERC),\\
	National Institute of Advanced Industrial Science and Technology (AIST),\\
	Tsukuba 305-8562, Japan}
\affiliation{Department of Advanced Materials,
	University of Tokyo, Kashiwa 277-8581, Japan.}
\author{Y. Takahashi}
\affiliation{Correlated Electron Research Center (CERC),\\
	National Institute of Advanced Industrial Science and Technology (AIST),\\
	Tsukuba 305-8562, Japan}
\author{T. Hasegawa}
\affiliation{Correlated Electron Research Center (CERC),\\
	National Institute of Advanced Industrial Science and Technology (AIST),\\
	Tsukuba 305-8562, Japan}
\author{Y. Tokura}
\affiliation{Correlated Electron Research Center (CERC),\\
	National Institute of Advanced Industrial Science and Technology (AIST),\\
	Tsukuba 305-8562, Japan}
\affiliation{Department of Applied Physics, University of Tokyo, Tokyo 113-8656, Japan}
\date{\today}
\begin{abstract}
Transport properties of SrTiO$_3$-channel field-effect
transistors with parylene organic gate insulator have been
investigated.
By applying gate voltage, the sheet resistance falls below
$R_{\Box}$\,$\sim$\,10\,k$\Omega$
at low temperatures, with carrier mobility exceeding
1000\,cm$^2$/Vs.
The temperature dependence of the sheet resistance taken under
constant gate voltage exhibits metallic behavior
($dR$/$dT$\,$>$\,0).
Our results demonstrate an insulator to metal transition in
SrTiO$_3$ driven by electrostatic carrier density control.
\end{abstract}
\maketitle
Electronic properties of transition-metal oxides, including
high-temperature superconductivity in cuprates and colossal
magnetoresistance in manganites, are currently an area of
active research\,\cite{Imada}.
These exotic phases are crucially dependent on 
the density of carriers in the system; for example, superconductivity 
in cuprates appears only at a certain range of carrier concentration.
Electrostatic carrier doping is an attractive approach in
this research field, because unlike chemical doping of
carriers ---a conventional method introducing lattice
disorder--- it enables us to tune only the carrier concentration
of materials under investigation\,\cite{Ahn,Inoue}.
Among transition-metal oxides, here we focus on SrTiO$_3$, a wide-gap insulator
(E$_g$\,=\,3.2\,eV) with a cubic-perovskite structure.
By chemical doping, SrTiO$_3$ becomes metallic, and by further
doping becomes a superconductor.
Interestingly, these transitions are reported to occur at low
carrier concentrations of
10$^{18}$\,cm$^{-3}$
and
10$^{19}$\,cm$^{-3}$,
respectively\,\cite{Tufte,Koonce}.
Since they are within a range accessible by electrostatic carrier
doping, the two phases may be observed by electrostatic carrier doping of
an undoped, insulating sample.

Recent attempts\,\cite{Ueno,Shibuya} to use undoped SrTiO$_3$
as the active channel of a field-effect transistor (FET)
have faced a common problem: carrier injection from
the source/drain electrodes becomes increasingly difficult
with decreasing temperature.
Even with a large applied gate electric field, the channel resistance
increases monotonically as temperature decreases\,\cite{Ueno}.
The origin may be trap levels in the SrTiO$_3$ channel, and/or barrier
formation at the interface between the metal electrodes and the
SrTiO$_3$ channel.

In this work, we attempted to realize a low temperature metallic state
in SrTiO$_3$-FET by solving this problem.
Our strategy is as follows:
First, we improve the quality of electrodes to improve
the metal-electrode/SrTiO$_3$-channel interface.
Next, we adopt an organic polymer, parylene, for a gate insulator to
minimize traps at the gate-insulator/SrTiO$_3$-channel interface.
Finally, we introduce two potential probes just inside the channel  
region to perform a four-probe measurement to determine the intrinsic channel
properties.
In this paper, we describe successful carrier injection at
low temperatures
using these strategies, 
and demonstrate an electric-field-induced
insulator-to-metal transition on an undoped SrTiO$_3$ single crystal.

\begin{figure}[!tb]
	\centering
	\includegraphics[width=6cm,clip]{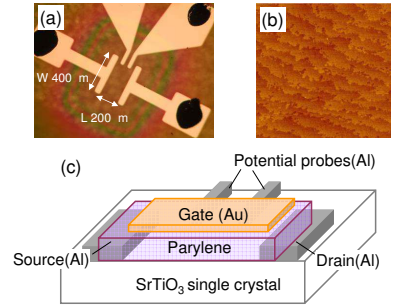}
	\caption{
(a) Photograph of our SrTiO$_3$-FET device taken just after the  
deposition of the parylene gate insulator.
(b) An image of the pristine SrTiO$_3$ (100) surface taken by
atomic force microscopy before device fabrication.
Displayed area is 2\,$\mu$m$\times$2\,$\mu$m.
(c) Schematic structure of SrTiO$_3$-FET.
Thickness of the aluminum electrodes is 20 nm and that of gate
electrode is 40 nm.
	}
\end{figure}%

The device structure is shown in Fig.\,1, which  
was
formed on the (100) surface of undoped SrTiO$_3$ single crystals.
(The crystals had been polished and etched by a vendor following 
Ref.\,\onlinecite{Kawasaki}.)
The 20\,nm thick aluminum electrodes were evaporated by
resistive heating using a tungsten boat under a pressure of
$10^{-4}$\,Pa.
Next, the parylene insulator ($\epsilon_r=3.15$) was deposited by first pyrolyzing the
monomer at 700$^\circ$C, and the polymerization on the SrTiO$_3$
substrate held at room temperature.
It should be noted that the deposition conditions, for aluminum
and parylene, markedly affects the device performance at low
temperatures, although many of the devices show almost identical
characteristics at room temperature.
We found that the key issue is the degree of vacuum during the deposition
of parylene.
The base pressure in which we deposit the parylene
must be kept lower than $3\times10^{-4}$\,Pa to prevent
the oxidization of aluminum, although the deposition of the
parylene itself does not in general require such high vacuum.
Finally, gold was evaporated for a gate electrode.
All the transport measurements were carried out by Agilent
E5287A high-resolution source/monitor unit modules
equipped on an E5270B.
Variable temperature was obtained by using a Quantum Design
physical property measurement system.

\begin{figure}[!tb]
	\centering
	\vspace{-0.7\intextsep}
	\includegraphics[width=6.3cm,clip]{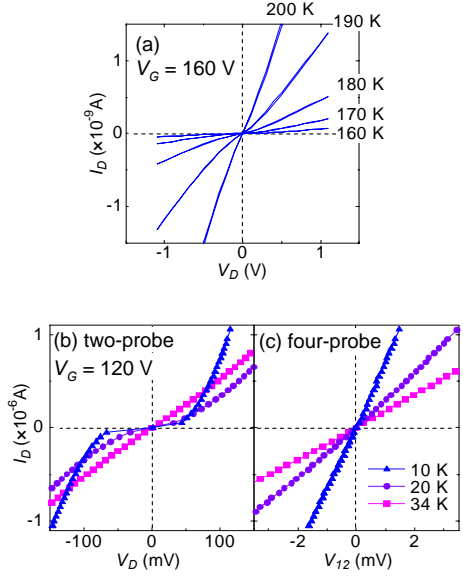}
	\caption{%
Current-voltage characteristics of parylene/SrTiO$_3$-FETs at low
temperatures.
(a)Typical drain current--drain voltage ($I_D$--$V_D$)
characteristic for a device with a parylene gate insulator
deposited in low vacuum (10$^{-3}$--10$^{-2}$\,Pa).
(b) $I_D$--$V_D$ curves for a device with a parylene gate
insulator deposited at higher vacuum (10$^{-4}$\,Pa).
(c) Same sample as (b) where $I_D$ is plotted against the voltage difference 
$V_{12}$ between the two potential probes inside the channel.
The parylene thickness was 0.53\,$\mu$m for both of devices.%
	}
\end{figure}%

As noted above, our SrTiO$_3$-FET with a parylene gate insulator  
deposited in low vacuum has poor carrier injection from
the source/drain electrodes at low temperature.
Figure\,2(a) shows typical drain current--drain  
voltage ($I_D$--$V_D$) characteristics of the device prepared at
low vacuum, where $I_D$ goes below the noise level below around 150\,K.
Figure\,2(b) shows $I_D$--$V_D$ curves of the device prepared at
higher vacuum, indicating the successful carrier injection at
low temperatures.
Nonlinear $I_D$--$V_D$ characteristics are still observed in the
small $V_D$ region.
Fig.\,2(c) shows those of
$I_D$--$V_{12}$, where $V_{12}$ is the potential drop inside
the channel measured by the two potential probes.
Ohmic behavior in Fig\,2(c) clearly indicate barrier formation at
the SrTiO$_3$/Al interfaces, which prevents carrier injection and
opens a gap in the small $V_D$ region in Fig\,2(b).
It is necessary to lower the barrier at the Al/SrTiO$_3$
interface for efficient carrier injection at low temperatures.

\begin{figure}[!tb]
	\centering
	\vspace{-0.7\intextsep}
	\includegraphics[width=6.3cm,clip]{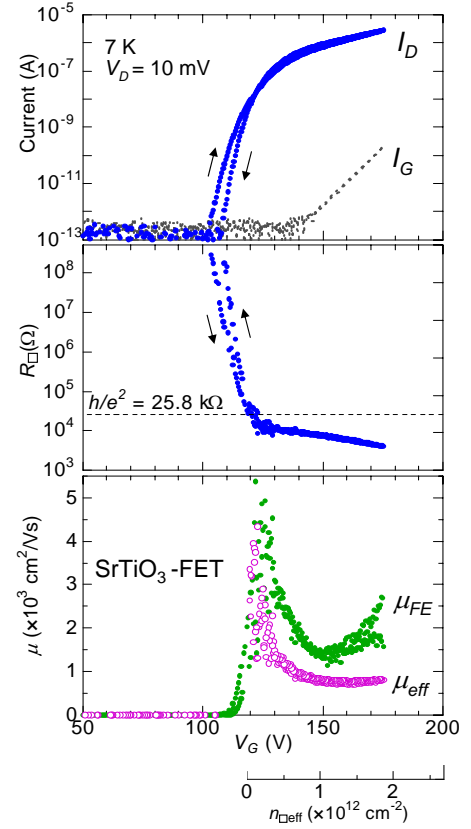}
	\caption{%
The drain current, $I_D$, the gate leakage current, $I_G$,
the sheet resistance, $R_\Box$, and the mobility, $\mu$,
are plotted against the gate voltage, $V_G$.
The measurement was carried out under constant drain
voltage of 10 mV.
The arrows indicate the sweep direction of the gate voltage,
and the observed hysterisis is probably due to traps at the
parylene/SrTiO$_3$ interface.
Sheet resistance was defined as
$R_{\Box}=(W/D)\cdot (I_{D}/V_{12})$,
where $V_{12}$ is the voltage drop between two potential probes,
$D$ is the distance between the two probes, and $W$ is the
channel width.	
A scale of the effective (mobile) electron density,
$n_{\Box\mathrm{eff}}$, is also shown at the bottom.
	}
	\label{fig7K}
\end{figure}

Now we turn to the details of the low-temperature characteristics
of the SrTiO$_3$-FET.
Figure\,3 shows the sheet resistance $R_ 
{\Box}$
as a function of the gate voltage $V_G$ at 8\,K.
Gate leakage current $I_G$ is much smaller than $I_D$, as also
plotted in Fig.\,3, which assures the accuracy  
of the measured sheet resistance.
As $V_G$ is increased, the effective sheet carrier density
$n_{\Box\mathrm{eff}}$\,=\,$C_i(V_G-V_ 
{G\mathrm{th}})$
is increased.
Here, $V_{G\mathrm{th}}$ is the threshold voltage of $\sim$118\,V
determined from a linear fit of the
$\sqrt{I_D}$--$V_G$
curve, and
$C_i$\,=\,5.3\,nF/cm$^2$
is the capacitance per unit area of the parylene gate insulator.
With increasing $n_{\Box\mathrm{eff}}$, $R_{\Box}$ decreases
drastically from more than G$\Omega$ to the order of k$\Omega$.

There are two points to be noted in the middle panel of Fig.\,3.
First, $R_{\Box}$ becomes smaller than the quantum 
resistance, $h/e^2$\,$\sim$\,25.8\,k$\Omega$.
Therefore, the channel region at higher gate voltages
can be metallic.
A rough estimate of the thickness of the conducting layer in the 
SrTiO$_3$-FET is less than 10 nm\,\cite{Inoue}.
Recent studies on two dimensional electron systems such as Si-MOSFET
or p-GaAs showed that a metal-insulator transition occurs when the
sheet resistance is the order of the quantum resistance\,\cite{Kravchenko}.
The second point is the presence of an ``inflection point'' in the
$R_{\Box}$--$V_G$ curve around 120\,V.
This point seems to reflect a qualitative change in the electronic
state of the channel.
Indeed, the value of $R_{\Box}$ at the inflection point is
almost identical to the quantum sheet resistance, underpinning
the insulator-to-metal transition in this system.

The large threshold $V_{G\mathrm{th}}$ observed in Fig.\,3
may be due to a high density of traps ($\sim$\,$10^{12}$\,cm$^{-2}$)
located at the SrTiO$_3$ surface.
Only when we apply sufficient gate voltage to fill the traps,
mobile electrons accumulate.
To estimate the mobility of the itinerant 
carriers, we consider two alternative expressions:
field-effect mobility, $\mu_{\mathrm{FE}}$, and the
effective mobility, $\mu_{\mathrm{eff}}$.
They are defined as follows:
\begin{align}
\mu_{\mathrm{FE}} &=\frac{\partial}{q 
\partial n_{\Box}}\left(\frac{1}{R_ 
\Box}\right)
                     =\frac{1}{C_i}\frac{\partial} 
{\partial V_G}\left(\frac{1}{R_\Box} 
\right) \notag\\
\mu_{\mathrm{eff}}&=\frac{1}{qn_{\Box  
\mathrm{eff}}R_{\Box}} \notag
\end{align}
The difference is that $\mu_{\mathrm{eff}}$ needs input of
the effective sheet carrier density
$n_{\Box \mathrm{eff}}$,
whereas $\mu_{\mathrm{FE}}$ does not.
The calculated mobilities are shown in the bottom panel of
Fig.\,3.
There are no significant difference between the two mobilities.
They are as high as 1000--2000\,cm$^2$/Vs, which is the highest 
mobility ever attained in an SrTiO$_3$-based FET\cite{Ueno,Shibuya,Pallechi,Pan,Takahashi}.
The large mobility supports our inference that a metallic state is  
induced by a large gate electric field at the surface of SrTiO$_3$ 
at low  temperatures.
To further confirm this, the temperature dependence of
sheet resistance has been measured.
Figure\,4 shows $R_{\Box}(T)$ curves  
taken at eight different
gate voltages; the larger gate voltage corresponds to the larger  
carrier density.
Apparently, metallic behavior ($dR_{\Box}$/$dT$\,$> 
$\,0) is observed
down to 8\,K, which is the lowest accessible temperature of the present  
measurement.

\begin{figure}[!t]
	\centering
	\includegraphics[width=6.3cm,clip]{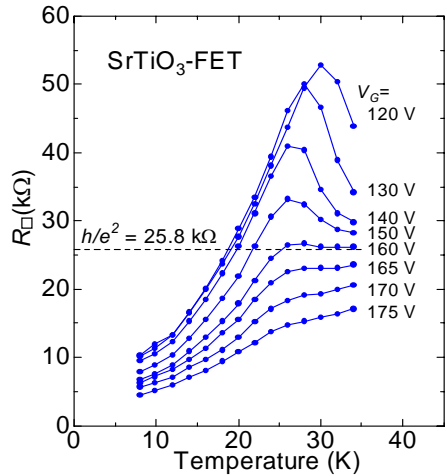}
	\caption{%
Sheet resistance $R_{\Box}$ as a function of temperature for
gate voltages $V_G$ of 120--175\,V.
The parylene gate insulator thickness is 0.53\,$\mu$m.
	}
	\label{figRs}
\end{figure}

It is interesting to note the presence of the anomaly around 25\,K
in Fig.\,4.
This anomaly becomes more prominent with decreasing carrier
density until a peak is formed in the $R_{\Box}(T)$ curve.
Above the characteristic temperature of the $R_{\Box}(T)$  
anomaly (25\,K--30\,K),
the $R_{\Box}(T)$ curves can be classified into
two groups: insulating behavior
   ($dR_{\Box}$/$dT$\,$>$\,0)
and metallic behavior
   ($dR_{\Box}$/$dT$\,$<$\,0).
Interestingly, at the border between these two behaviors, where the
plateau in $R_{\Box}(T)$ curve is observed ($V_G$\, 
$\sim$\,160\,V), the sheet
resistance is very close to $h/e^2$.

In summary, improvement of the fabrication process of
single-crystalline SrTiO$_3$-based FET, has led to the finding
that the surface of the undoped SrTiO$_3$ turns into a high-mobility
($>$1000\,cm$^2$/Vs) conducting state by applying a large  
gate electric field at low temperatures.
Temperature dependence of the sheet resistance $R_{\Box}$
suggests the high-mobility state is indeed metallic ($dR_ 
{\Box}$/$dT$\,$>$\,0), with $R_{\Box}$ below the quantum resistance for
$V_G$\,$>$\,160\,V.
Field-induced superconductivity of undoped SrTiO$_3$
is our current target for further deployment of this technique.

This work is partially supported by Grant-in-Aid for Scientific 
Research from MEXT, Japan.
We thank H. Y. Hwang for discussion and critical reading of the manuscript.

\end{document}